# Dependence of tunnel magnetoresistance on ferromagnetic electrode materials in MgO-barrier magnetic tunnel junctions

Shoji Ikeda[a,*], Jun Hayakawa[b,a], Young Min Lee[a], Fumihiro Matsukura[a], Hideo Ohno[a]

[a] *Laboratory for Nanoelectronics and Spintronics, RIEC, Tohoku University, 2-1-1 Katahira, Aoba-ku, Sendai 980-8577, Japan*

[b] *Advanced Research Laboratory, Hitachi, Ltd., 1-280 Higashi-koigakubo, Kokubunji-shi, Tokyo 185-8601 ,Japan*

## Abstract

We investigated the relationship between the tunnel magnetoresistance (TMR) ratio and the electrode structure in MgO-barrier magnetic tunnel junctions (MTJs). The TMR ratio in a MTJ with $Co_{40}Fe_{40}B_{20}$ reference and free layers reached 355% at the post-deposition annealing temperature of $T_a$=400°C. When $Co_{50}Fe_{50}$ or $Co_{90}Fe_{10}$ is used for the reference layer material, no high TMR ratio was observed. The key to have high TMR ratio is to have highly oriented (001) MgO barrier/CoFeB crystalline electrodes. The highest TMR ratio obtained so far is 450% at $T_a$ = 450°C in a pseudo spin-valve MTJ.

* Corresponding author: sikeda@riec.tohoku.ac.jp.

To realize high-speed magnetoresistive random access memory, it is strongly desired that magnetic tunnel junctions (MTJs) that constitute memory cells show high tunnel magnetoresistance (TMR). Developments of MTJs with highly (001)-oriented MgO barrier/ferromagnetic electrodes in recent years, motivated by pioneering theoretical predictions [1,2], resulted in very high TMR ratio ranging from 180 % to 355% at room temperature (RT) [3-9]. However, the details of ferromagnetic electrodes which yield giant TMR ratio have not yet been fully investigated. In this work, we studied the relationship between the TMR ratio and the film structure in MgO-barrier MTJs with different ferromagnetic electrodes.

The MTJ multilayer structure studied here was Si/$SiO_2$ wafer/ Ta(5)/ Ru(50)/ Ta(5)/ NiFe(5)/ MnIr(8)/ CoFe(2)/ Ru(0.8)/ reference layer(3)/ MgO(1.5)/ free layer(3)/ Ta(5)/ Ru(15) (in nm) deposited using RF magnetron sputtering, where $Co_{40}Fe_{40}B_{20}$, $Co_{50}Fe_{50}$, and $Co_{90}Fe_{10}$ alloys (in at%) were used for the reference and free layers. In addition, we studied a pseudo-spin valve structure of Si/$SiO_2$ wafer/ Ta(5)/ Ru(50)/ Ta(5)/ $Co_{40}Fe_{40}B_{20}$(3)/ MgO(1.7)/ $Co_{40}Fe_{40}B_{20}$(4)/ Ta(5)/ Ru(15). All MTJs were micro-fabricated by photolithography with a junction size of 0.8×4.0 $\mu m^2$, and were annealed at 250 to 425-450 $^{o}C$ in a vacuum under a field of 4 kOe. We measured TMR using a dc four-probe method at RT in the magnetic field range of ±1 kOe, and defined the TMR ratio as $(R_{ap}-R_p)/R_p \times 100$, where $R_p$ and $R_{ap}$ are the resistance at parallel (P) and antiparallel (AP) configurations between the reference and free layers, respectively. The film structure was characterized by x-ray diffraction (XRD) using Cu K$\alpha$ radiation.

Figure 1 shows TMR ratio as a function of the annealing temperature ($T_a$) for the MTJs with $Co_{40}Fe_{40}B_{20}$, $Co_{50}Fe_{50}$, and $Co_{90}Fe_{10}$ reference layers, in which the free layer was fixed to $Co_{40}Fe_{40}B_{20}$. The TMR ratio in the MTJ with the $Co_{40}Fe_{40}B_{20}$ reference layer increases with increasing $T_a$ and reaches 355% at $T_a$=400$^{o}$C. When $Co_{50}Fe_{50}$ or $Co_{90}Fe_{10}$ was used for the reference layer material, no high TMR ratio was observed. For the MTJs with $Co_{50}Fe_{50}$ and $Co_{90}Fe_{10}$ reference layers, the maximum TMR ratio was 50% and 75%, respectively.

To understand the difference of the TMR ratios for the MTJs with three different reference layers, the crystalline structures were examined by XRD on as-deposited samples prepared for the XRD measurements. As-deposited $Co_{40}Fe_{40}B_{20}$ has amorphous structure. As can be seen in Fig. 2(a), the formation of highly (001)-oriented MgO on the $Co_{40}Fe_{40}B_{20}$ amorphous reference layer is observed. The MgO layer acts as a template for crystallization to highly (001)-oriented $Co_{40}Fe_{40}B_{20}$ through annealing at high temperatures. The $Co_{50}Fe_{50}$ and $Co_{90}Fe_{10}$ layers are bcc and fcc polycrystalline, respectively. Although MgO on the $Co_{50}Fe_{50}$ reference layer has a (001)-oriented structure (Fig.2(b)), the (001) degree of orientation on the $Co_{50}Fe_{50}$ reference layer is low compared with that on the $Co_{40}Fe_{40}B_{20}$. MgO on the $Co_{90}Fe_{10}$ reference layer has fcc(111)-oriented texture (Fig.2(c)), showing that the structure of reference layer strongly affects the structure of MgO. These structural characteristics are qualitatively consistent with the results of the TMR ratio in Fig. 1. Highly oriented (001) MgO barrier/ $Co_{40}Fe_{40}B_{20}$ crystalline electrodes result in high TMR ratio of 355%. In contrast, the MTJs with the polycrystalline bcc-$Co_{50}Fe_{50}$ and fcc $Co_{90}Fe_{10}$ reference layers did not show high TMR ratio, which can be attributed to the absence of highly (001)-oriented MgO barrier/ferromagneric electrodes, where $\Delta_1$ bands in bcc (001) electrodes and its selective tunneling in MgO (001) barrier responsible for the high TMR are absent.

Table I shows the maximum TMR ratio and the resistance- area product (*RA*) in parallel configuration for the 1.5 nm thick MgO-barrier MTJs with different reference and free layers annealed at optimum temperature ($T_a^{op}$). The application of the $Co_{40}Fe_{40}B_{20}$ reference layer to the MTJs results in a high thermal stability, i.e., high TMR ratio through annealing at high $T_a$. No clear correlation between the TMR ratio and *RA* has been seen.

The TMR ratio depends on the MTJ stack structure besides the reference and free layers which are in direct contact with the MgO barrier. For example, the MTJ stack composed of CoFe/CoFeB pinned layer without Ru spacer exhibited a low TMR ratio of 181% at $T_a$ = 325°C, because CoFeB on CoFe crystallized to a bcc (110) structure by annealing [10, 11]. A pseudo spin-valve MTJ, in which the CoFeB layers are not influenced by the layers in contact with them even at high $T_a$, showed a high TMR ratio of 450% at $T_a$ = 450°C.

This work was supported by the IT-program of Research Revolution 2002 (RR2002): “Development of Universal Low-power Spin Memory” from the Ministry of Education, Culture, Sports, Science and Technology of Japan.

Figure captions

Fig.1 Annealing temperature dependence of TMR ratio for the MTJs with $Co_{40}Fe_{40}B_{20}$, $Co_{50}Fe_{50}$ and $Co_{90}Fe_{10}$ reference layers.

Fig. 2 XRD patterns for the as-deposited samples consisting of substrate/ Ta(5) /*X*(3) / MgO(50) with *X* = (a) $Co_{40}Fe_{40}B_{20}$, (b) $Co_{50}Fe_{50}$, and (c) $Co_{90}Fe_{10}$.

Table I Maximum TMR ratio and *RA* for the MTJs with different reference and free layers annealed at optimum temperature ($T_a^{op}$) that gives the maximum TMR ratio.

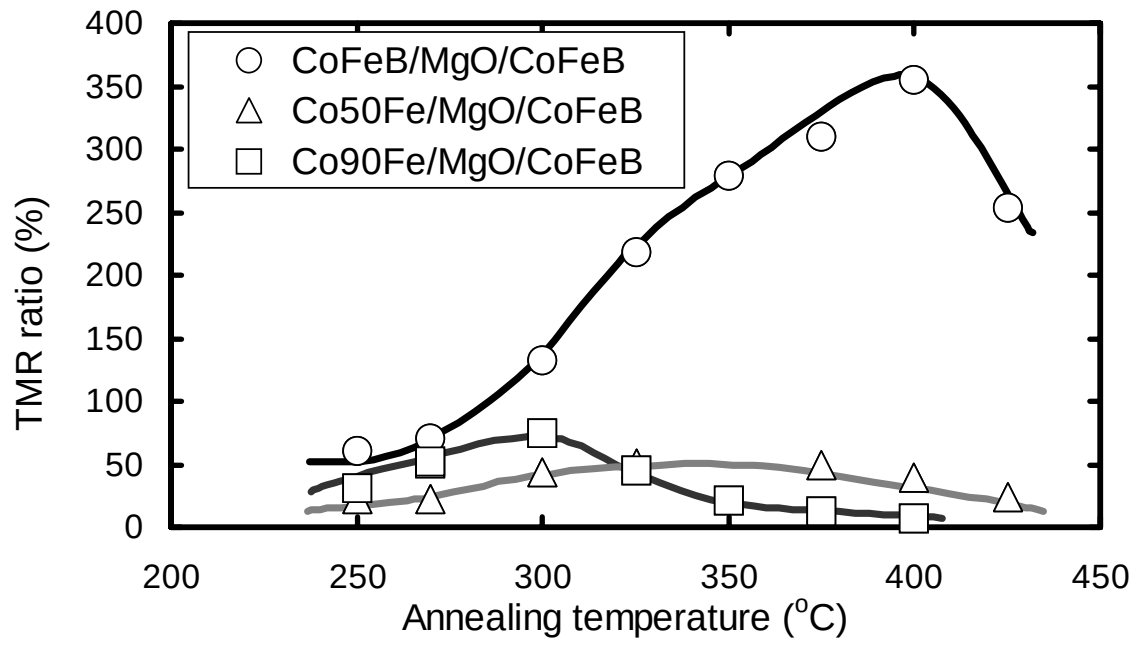

Fig. 1

S. Ikeda et al.

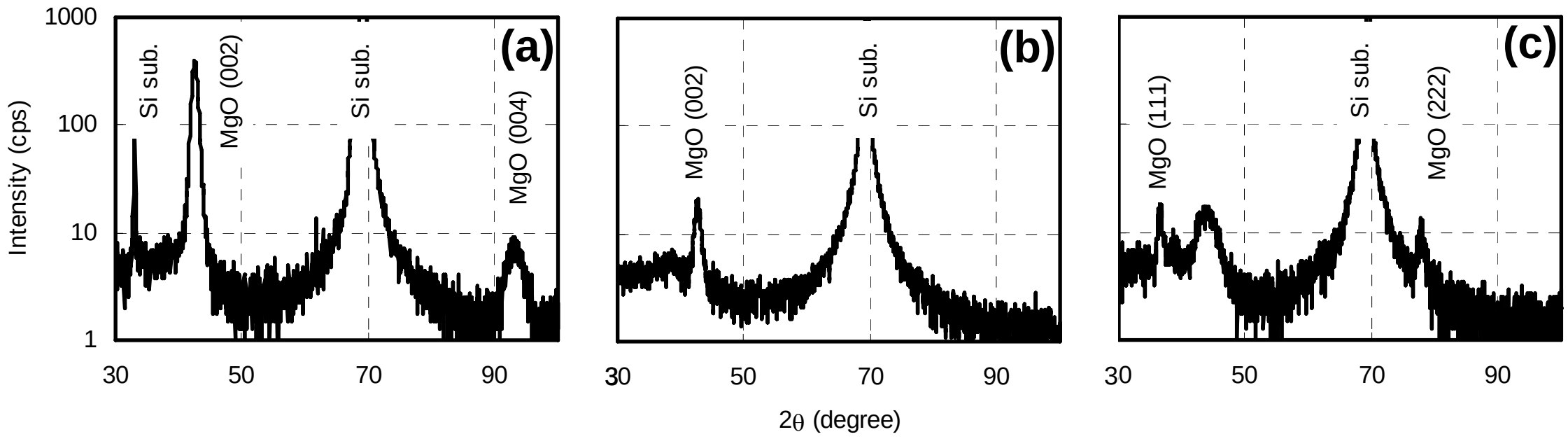

Fig. 2

S. Ikeda et al.

Table I

| Reference layer | Free layer | $T_a^{op}$ ($^oC$) | TMR ratio (%) | *RA* ($\Omega\mu m^2$) |
|---|---|---|---|---|
| $Co_{40}Fe_{40}B_{20}$ | $Co_{40}Fe_{40}B_{20}$ | 400 | 355 | 547 |
| $Co_{40}Fe_{40}B_{20}$ | $Co_{50}Fe_{50}$ | 400 | 277 | 1060 |
| $Co_{40}Fe_{40}B_{20}$ | $Co_{90}Fe_{10}$ | 350 | 131 | 714 |
| $Co_{50}Fe_{50}$ | $Co_{40}Fe_{40}B_{20}$ | 325 | 50 | 1042 |
| $Co_{50}Fe_{50}$ | $Co_{50}Fe_{50}$ | 270 | 12 | 740 |
| $Co_{90}Fe_{10}$ | $Co_{40}Fe_{40}B_{20}$ | 300 | 75 | 475 |
| $Co_{90}Fe_{10}$ | $Co_{90}Fe_{10}$ | 270 | 53 | 571 |
| (pseudo-spin valve with 1.7 nm thick MgO-barrier) | | | | |
| $Co_{40}Fe_{40}B_{20}$ | $Co_{40}Fe_{40}B_{20}$ | 450 | 450 | 3700 |

S. Ikeda et al.